\documentclass[twocolumn,amsmath,amssymb,superscriptaddress,floatfix]{revtex4}
\usepackage{color}
\usepackage{epsfig}
\usepackage{epstopdf}
\usepackage{grffile}
\usepackage{bm}
\usepackage{sidecap}
\usepackage{mathptmx}
\usepackage{txfonts}
\usepackage[colorlinks,citecolor=blue,linkcolor=blue,urlcolor=blue]{hyperref}
\usepackage[usenames,dvipsnames,svgnames]{xcolor}
\usepackage{array}
\usepackage{float}

\usepackage{lipsum}
\usepackage{graphicx}
\usepackage{dcolumn}
\usepackage{bm}
\usepackage{tabularx}
\usepackage{tabu}
\usepackage{multirow}
\usepackage{array}
\usepackage{booktabs}
\usepackage{tabularx} 
\usepackage{ulem}

\begin{document}

\title{Impact of strain on the excitonic linewidth in transition metal dichalcogenides} 
\author{Zahra Khatibi}
\email{za.khatibi@gmail.com}
\affiliation{Chalmers University of Technology, Department of Physics, 412 96 Gothenburg, Sweden}
\affiliation{Iran University of Science and Technology, Department of Physics,  Narmak, 16846-13114, Tehran, Iran}
\author{Maja Feierabend}
\affiliation{Chalmers University of Technology, Department of Physics, 412 96 Gothenburg, Sweden}
\author{Malte Selig}
\affiliation{Institut f\"ur Theoretische Physik, Technische Universit\"at Berlin, 10623 Berlin, Germany}
\affiliation{Chalmers University of Technology, Department of Physics, 412 96 Gothenburg, Sweden}
\author{Samuel Brem}
\affiliation{Chalmers University of Technology, Department of Physics, 412 96 Gothenburg, Sweden}
\author{Christopher Linder\"alv}
\affiliation{Chalmers University of Technology, Department of Physics, 412 96 Gothenburg, Sweden}
\author{Paul Erhart}
\affiliation{Chalmers University of Technology, Department of Physics, 412 96 Gothenburg, Sweden}
\author{Ermin Malic}
\affiliation{Chalmers University of Technology, Department of Physics, 412 96 Gothenburg, Sweden}

\begin{abstract}
Monolayer transition metal dichalcogenides (TMDs) are known to be highly sensitive to externally applied tensile or compressive strain. In particular,  strain can be exploited as a tool to control the optical response of TMDs. However, the role of excitonic effects under strain has not been fully understood yet. Utilizing the strain-induced modification of electron and phonon dispersion obtained by first principle calculations, we present in this work microscopic insights into the strain-dependent optical response of various TMD materials. In particular, we explain recent experiments on the change of excitonic linewidths in strained TMDs and predict their behavior for tensile and compressive strain at low temperatures. 
\end{abstract}
\maketitle

\section{Introduction}
\begin{figure}[b!]
\includegraphics[width= 0.75\linewidth]{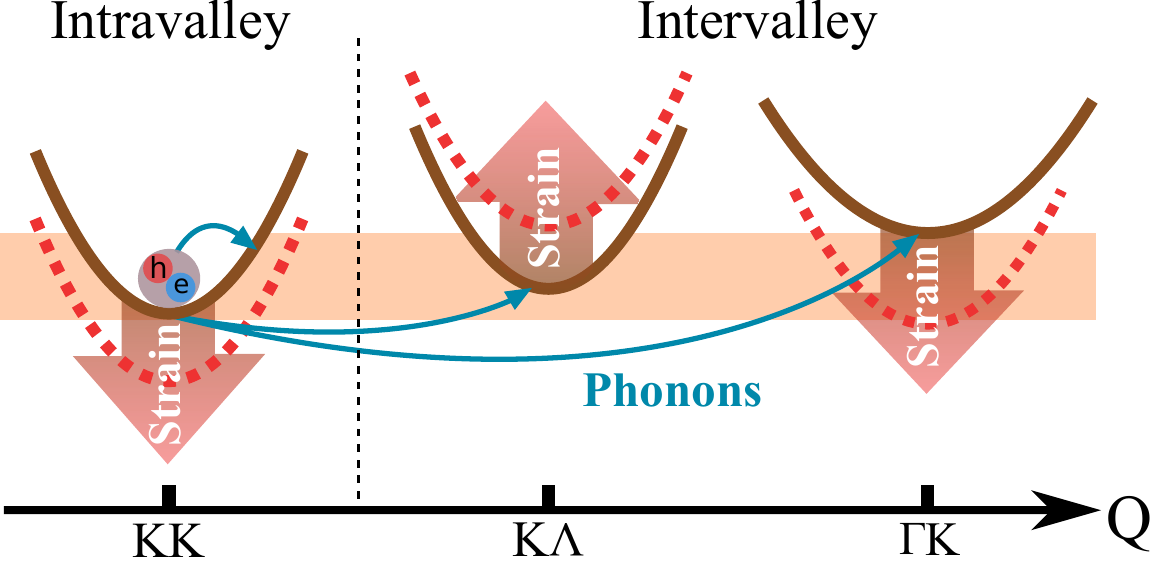}
\caption{
Schematic representation of exciton-phonon processes in the exemplary $\rm{MoS_2}$ monolayer. 
The dashed red lines indicate excitonic bands in presence of a tensile strain. 
Intra- and intervalley phonon-assisted scattering channels are demonstrated via blue arrows, while the strain-induced shift of excitonic bands is denoted by red arrows.
To aid visualization, the energetic separation of the bright KK and the momentum-forbidden dark K$\Lambda$ and $\Gamma$K excitonic states is denoted by the orange stripe.
\label{fig1}}
\end{figure}

Monolayer transition metal dichalcogenides (TMDs) exhibit an efficient light-matter interaction resulting in an pronounced absorbance and photoluminescence  \cite{mak2010atomically,splendiani2010emerging,tonndorf2013photoluminescence}. Due to their atomically thin structure, they are remarkably sensitive to changes in the environment \cite{schmidt2016reversible,island2016precise}. 
Mechanical deformations imposed on their lattice structure drastically affect their optical and electronic properties \cite{yue2012mechanical,Johari2012,horzum2013phonon,steinhoff2014influence,wang2014many,Riccardo2017,Gorbani2018}.
In particular, significant modifications of the optical response have been observed \cite{he2013experimental,castellanos2013local,plechinger2015control,Iris2018,Xie2018,Mennel2018}. Exciton resonances shift towards red (blue) when applying tensile (compressive) strain, which is well understood and is due to a decrease (increase) of the electronic band gap (Fig. \ref{fig1}). 
Furthermore,  the width of the excitonic resonance changes considerably, when strain is applied.
Surprisingly, different TMD materials behave contrary with respect to strain. 
While the exciton linewidths become broader in MoS$_{2}$, WSe$_{2}$ shows a narrowing of the linewidth, when the material is stretched \cite{Iris2018}. 
These observations are not yet fully understood and to exploit the promising potential of TMDs for strain engineering, it is necessary to gain microscopic insights into the underlying fundamental processes.

In this work, we investigate strain-induced changes in the electronic and phonon dispersion by utilizing first principle calculations. We exploit this knowledge as an input for the solution of microscopically derived semiconductor Bloch equations providing access to the phonon-induced excitonic line broadening. Including all relevant intra- and intervalley scattering channels (Fig. \ref{fig1}), we find that the strain-induced changes in the linewidth can be clearly traced back to the change in the exciton landscape, i.e. the relative energetic positions of different excitonic states. The theoretically predicted changes in the linewidth show an excellent agreement with recent experiments \cite{Iris2018} and predict new features for yet not realized low-temperature and compressive strain measurements.
\section{Theory}

When a TMD is exposed to an external in-plane force, it can be either stretched or compressed out of its equilibrium state. Biaxial tensile (compressive) strain expands (compresses) the lattice in both in-plane directions equally and restores the lattice symmetries, while uniaxial strain distorts the lattice in only one in-plane direction breaking the original lattice symmetry.
Consequently, both affect the electronic band structure through the lattice distortion and a change in the overlap of the atomic wave functions \cite{Chang2013,Maja2017}.
Furthermore, the displacement of the atoms out of their equilibrium positions leads to changes in the phonon modes, which are crucial for understanding exciton-phonon scattering channels in different TMD materials \cite{Malte2016}.

We study the strain-dependent changes in the optical response of different TMD materials by combining  DFT calculations of the electronic and phonon dispersions with semiconductor Bloch equations.
First-principle calculations are performed within the \textsc{Quantum} ESPRESSO package \cite{qe2009}.
We employ plane wave functions with ultrasoft pseudopotential alongside the LDA exchange-correlation functional within the Davidson method \cite{qe2009}. We apply 
$18\times 18 \times 1$ Monkhorst--Pack (MP) k-point grids and 60 Ry mesh cutoff. 
All simulations include a vacuum space of approximately 20~\AA~height to exclude any interactions with spurious images of TMD layers. 
The phonon dispersion of the TMDs is calculated utilizing DFPT method within the Phonon package \cite{Baroni2001,qe2009}.
We sample the phonon momentum space with a $6\times 6 \times 1$ MP grid.
For both unstrained and strained TMDs we use the same convergence threshold, energy mesh cutoff, and the same configuration for the sampling of the momentum space.
%
%
\begin{figure}[!t]
\includegraphics[width= \linewidth]{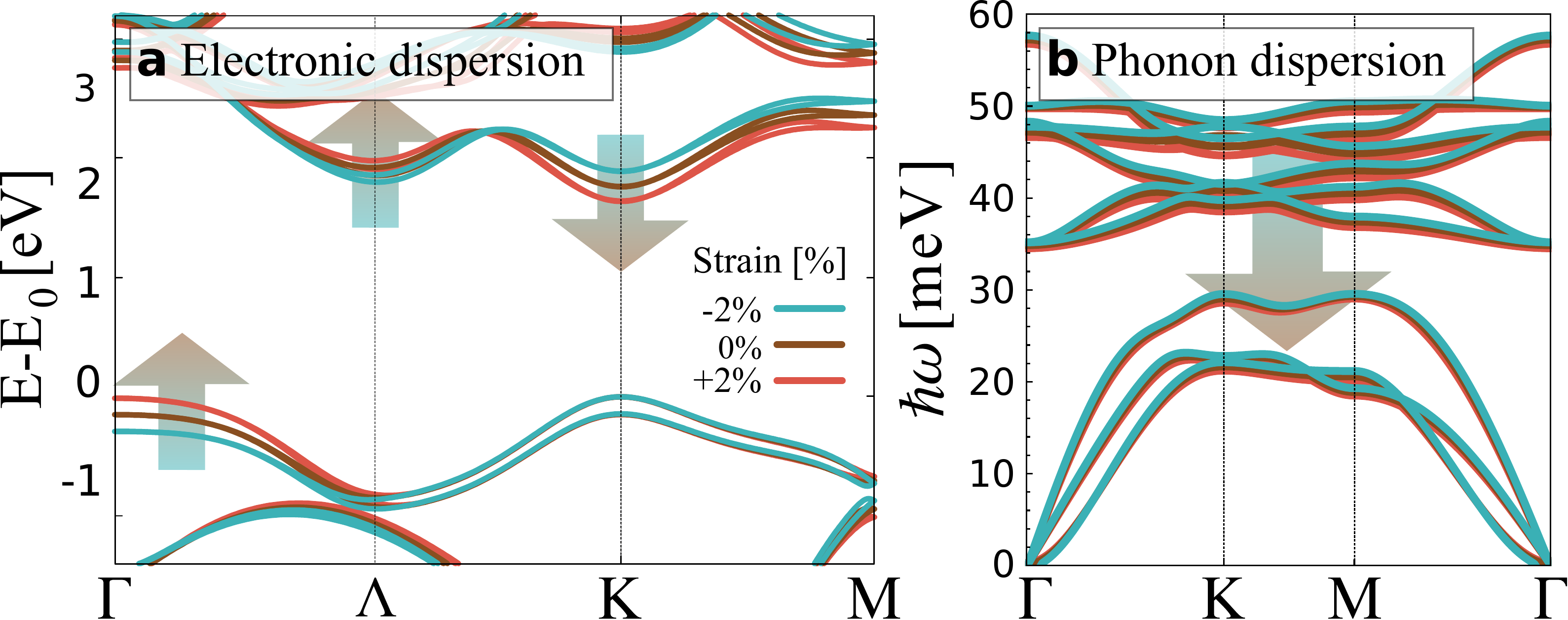}
\caption{(a) Electronic and (b) phonon dispersion for the exemplary ${\rm MoS_2}$ monolayer subjected to tensile and compressive strain.
The dispersions are plotted along the high symmetry points connecting the corner (K) to the center ($\Gamma$) of the Brillouin zone. 
The maximum of the valence band at the K valley is fixed at ${0~\rm eV}$.
Starting from the compressive strain and stretching ${\rm MoS_2}$ monolayer out, we observe an energetic shift of the K, $\Lambda$ and $\Gamma$ valleys. Interestingly, the K valley shifts downward with respect to the tensile strain, while
$\Gamma $ and $\Lambda$ valleys move to higher energies.
The change in the phonon dispersion is less pronounced exhibiting a slight decrease in energies in presence of strain. \label{fig2}}
\end{figure}

For an unstrained system, we relax the lattice structure, so that the forces on each atom become less than 0.001~$\frac{{\rm Ry}}{{\rm a.u.}}$.
To strain the TMDs within DFT method, the lattice vectors are modeled as,
$\vec{r}^{~\prime}_{i}=(1+\varepsilon_{i})\vec{r}_{i}$, 
in which the strain is denoted by $\varepsilon_{i}$, $i=x,y$ is the index for in-plane axes and $\vec{r}_{i}$ is the corresponding in-plane lattice vector.
Hence, biaxial tensile (compressive) strain is interpreted as the increase (decrease) of the lattice constant relative to its equilibrium magnitude. 
To obtain the electronic and phonon dispersion along the high symmetry points in the Brillouin zone, we relax the structure, while setting the lattice vectors to strained values using the same force threshold.
Therefore, only the atomic coordinates are let to be relaxed with regard to strained vectors. 

Figure \ref{fig2} shows the electronic and phonon dispersion for MoS$_2$ monolayer as an exemplary TMD along high symmetry points for zero, tensile and compressive strain. Here, we define the direct KK gap as the energetic difference between the valence band maxima and the conduction band minima with the same spin at the K valley. Also, note the indirect K$\Lambda$ ($\Gamma$K) gap denoted as $E_{{\rm K\Lambda}}$ ($E_{{\rm \Gamma K}}$) is the energetic separation of the valence band maxima at the K ($\Gamma$) valley and the conduction band minima with the same spin at the $\Lambda$ (K) valley. Interestingly, we find that for all TMDs the direct KK and the indirect K$\Lambda$, $\Gamma$K gaps behave differently when being exposed to tensile strain. While $E_{\text{KK}}$ and $E_{{\rm \Gamma K}}$ decrease,  $E_{{\rm K\Lambda}}$ increases in presence of tensile strain, cf. the arrows in Fig. \ref{fig2}. 
For compressive strain, we find the opposite behavior. This behavior can be traced back to the different orbital composition of the involved bands \cite{Cappelluti2013, Rasmussen2015}.
 Note that actually, all bands show a down (up) shift for tensile (compressive) strain, however relative to the valence band, the conduction band at the K valley shifts stronger than that at the $\Lambda$ valley resulting in an opposite shift. Furthermore, we find that the energy shifts are linear in strain and show the same trends in all investigated TMDs. Although uniaxial strain distorts the lattice anisotropically, both uniaxial and biaxial strain has the same qualitative effects on the electronic dispersion. The biaxial strain is quantitatively roughly twice stronger than the uniaxial strain in terms of the observed spectral shifts. This implies that in the investigated strain range, the anisotropic nature of the uniaxial strain has only a minor effect on the overlap of orbital functions. Therefore, when we state in-plane strain we refer to uniaxial tensile strain unless stated otherwise. Furthermore, note that as the strain-induced modification of the splitting between different spin bands at all high symmetry points is negligibly small, the energetic shift of the K$\Lambda$ and K$\Lambda^{\prime}$ states is the same.

\begin{table}[b!] 
\centering
\begin{tabularx}{.48\textwidth}{X X X X X}
\hline
\hline
Electroinc&${\rm MoS_2}$  &  ${\rm MoSe_2}$  & ${\rm WS_2}$  &  ${\rm WSe_2}$\\
\hline
${\rm Unstrained}$  &  &  &  &   \\
\hline
$a_0$ & 3.13& 3.29& 3.15& 3.28\\
 $ m^{c,{\rm K}}$          & 0.42    & 0.52  & 0.26     &  0.32 \\ 
$ m^{c,\Lambda}$    & 0.58    & 0.55  & 0.5      &  0.45 \\ 
$ m^{v,{\rm K}}$              &-0.48    &-0.62  &-0.34    & -0.39    \\
$ m^{v,\Gamma}$      &-2.15    &-4.5      &-2.4    & -3.5     \\
$ E_{{\rm K \Lambda}}$      & 107    & 171      & 55    &  1     \\
$ E_{{\rm K \Lambda^{\prime}}}$      & 180    & 196     & 319    & 218     \\
$ E_{{\rm K \Gamma}}$      & 145     & 317      & 237    &  484     \\
\hline
${\rm Strained}$  &  &  &  &  \\
 \hline
 $\Delta m^{c,{\rm K}}$          & $\mp$0.012     & $\mp$ 0.013    &  $\mp$ 0.007  & $\mp$ 0.01 \\ 
$\Delta m^{c,\Lambda}$    & $\pm$ 0.008    & 0.0             &  0.0            & $\pm$0.005 \\ 
$\Delta m^{v,{\rm K}}$             & $\pm$0.008    & $\pm$0.005        &  $\pm$0.005     & $\pm$0.005 \\
$\Delta m^{v,\Gamma}$     & $\pm$0.33        & $\pm$0.8        &  $\pm$0.28     & $\pm$0.3\\
$\Delta E_{{\rm K \Lambda^{(\prime)}}}$     & $\pm 81$    & $\pm 76.5$         &  $\pm 95$     & $\pm 87$ \\
$\Delta E_{{\rm K \Gamma}}$     & $\mp 66.5$     & $\mp 61.5$     &  $\mp 64$     & $\mp 53$ \\
$\Delta E_{gap}$        & $\mp 73$     & $\mp 69$     &  $\mp 60$     & $\mp 53$\\
\hline
\hline
\end{tabularx}
\bigskip

\begin{tabularx}{.48\textwidth}{X X X X X}
\hline
\hline
Phonon&${\rm MoS_2}$  &  ${\rm MoSe_2}$  & ${\rm WS_2}$  &  ${\rm WSe_2}$\\
\hline
${\rm Unstrained}$  &  &  &  &  \\
\hline
$ E_{\Lambda}$  &  23.6 & 16.5 & 18.6 & 14.1\\
$E_{K}$ &  29.1 &  19.9 & 22.9 &  17.4  \\
\hline
${\rm Strained}$  &  &  &  &  \\
 \hline
$\Delta E_{\Lambda}$   & -0.038 & -0.041 & -0.050 & -0.034  \\
$\Delta E_{K}$ &  -0.067  & -0.042 & -0.057 & -0.045 \\
\hline
\hline
\end{tabularx}
\caption{Lattice, electronic and phonon dispersion parameters extracted from DFT calculations for all investigated TMDs in the unstrained and strained case. The strained values correspond to the absolute changes of the parameters per percent of tensile strain.
The strain effect on the electronic parameters is linear and hence can be calculated for any strain strength.
Energetic separations between the valleys are in units of meV.
Effective masses are in units of electron mass ($m_0$). 
The displayed phonon dispersion energies correspond to longitudinal acoustic (LA) phonons in units of meV. The subscript indicates the phonon wave vector.
\label{table1} 
}\end{table}
%
%
%
%

Despite the pronounced shifts in the electronic band structure in presence of strain, the phonon dispersion for all modes and wave vectors undergoes only a small decrease (less than 0.4\% per percent strain), cf. Fig. \ref{fig2}(b).
The set of electronic and lattice parameters obtained from DFT calculation is listed in Table \ref{table1}.
Our results are in good agreement with previous DFT studies regarding the changes in effective masses \cite{Hongliang2013,yue2012mechanical}, experimentally observed spectral shifts regarding the direct band gap \cite{Iris2018}, and experimental reports of the strain-induced phonon energy shifts \cite{Conley2013}.

Now, we use the calculated strained and unstrained electronic and phonon parameters as input for the solution of the semiconductor Bloch equations \cite{KIRA2006}.
The first step is to compute the excitonic band structure (Fig. \ref{fig1}) starting from the electronic dispersion (Fig. \ref{fig2} (a)). We approximate the latter parabolically in the vicinity of the high symmetry points and solve the Wannier equation \cite{Gunnar2014},
\begin{equation}
\frac{\hbar^2 \mathbf{q}^2}{2m^{\xi}}\varphi^{\mu \xi}_\mathbf{q} -\sum_\mathbf{k} V^{\xi}_\mathbf{k-q} \varphi^{\mu \xi}_\mathbf{k} = E_B^{\mu \xi} \varphi_\mathbf{q}^{\mu \xi}   
\label{Wannier}
\end{equation}
which provides access to exciton binding energies $E_B^{\mu \xi}$ and wavefunctions $\varphi_\mathbf{q}^{\mu \xi}$ for all exciton states $\mu$ including indirect excitons with electrons and holes located at different high symmetry points in the Brillouin zone \cite{Ermin2018}.
Here, $m^{\xi}$ denotes the reduced mass of the exciton with $\xi=(\xi_e,\xi_h)$ being a compound index for the electron (e) and hole (h) valley. We include all relevant minima in the conduction band with $\xi_e=K,K',\Lambda,\Lambda'$ and  all maxima of the valence band with $\xi_h=K,K',\Gamma$. 
Furthermore, $ V^{\xi}_\mathbf{k-q}$ denotes the Coulomb potential where we make use of the Rytova-Keldysh formalism for charges in thin films \cite{rytova1967screened,Keldysh,Timothy2013}. 
Throughout this work, we focus on the exciton ground state $A_{1s}$ which is the energetically lowest lying exciton with parallel electron and hole spins, i.e. $\mu=1s$. 

Having access to strain-dependent exciton binding energies, electron band gaps, and electronic effective masses, we approximate the excitonic dispersion as 
\begin{equation}
E^{\xi}_\mathbf{Q}=\sum_\mathbf{\xi} \left(E_0^{\xi}+E_B^{\xi}\right) +\frac{\hbar^2 (\mathbf{Q}-\mathbf{Q}_0^\xi)^2}{2 M^{\xi}}
\label{ex-dis}
\end{equation}
with $M^{\xi}$ denoting the total mass of the respective exciton, $E^{\xi}_0$ as the quasi-particle bandgap of the respective exciton and $\mathbf{Q}_0^{\xi}$ describing the momentum difference of the high symmetry points of electron and hole.
Finally, the excitonic 2D material Bloch equation reads in the linear limit \cite{Malte2016,Malte2018}
\begin{equation}
\partial_t P^{\xi}_\mathbf{Q} = - \frac{i}{\hbar} \left(E^{\xi}_\mathbf{Q} - i\frac{\gamma_\mathbf{Q}^{\xi}}{2}\right) P^{\xi}_\mathbf{Q} + \Omega^{\xi} \delta^{\xi_e,\xi_h}_{\mathbf{Q},0}.
\end{equation}
Here, the first term accounts for the oscillation of the excitonic coherence $P^{\xi}_\mathbf{Q}$ with its eigen frequency $E^{\xi}_\mathbf{Q}$ and the decay via the dephasing rate $\gamma_\mathbf{Q}^{\xi}$, which will be in the focus of this work. The last term describes the excitation of the excitonic polarization with an external light field, where the Rabi frequency  in the excitonic basis is defined as $\Omega^{\xi}= \sum_{\mathbf{q}} \varphi^{* \xi}_\mathbf{q} \mathbf{M}^{\xi}_\mathbf{q} \cdot \mathbf{A}$. Here $\mathbf{A}$ denotes the vector potential of the incident light field and $\mathbf{M}^{\xi}_\mathbf{q} = \langle c,\xi_e+\mathbf{q} |\nabla| v,\xi_h+\mathbf{q} \rangle$ the optical matrix element.
The dephasing of the excitonic coherence is determined by radiative and non-radiative contributions. The first is obtained from a self-consistent solution of Maxwell Bloch equations and has been calculated in analogy to Ref. \cite{Malte2016}. 

The non-radiative contribution to the dephasing is determined by exciton-phonon scattering in the low excitation limit. The corresponding matrix elements read \cite{Malte2016}
\begin{equation}
g^{\xi\xi'\nu}_\mathbf{q}=\sum_\mathbf{k} \varphi^{*\xi}_\mathbf{k} \left( g^{c \nu}_{\xi'-\xi+\mathbf{q}} \varphi^{\xi'}_\mathbf{k+\alpha q} - g^{v \nu}_{\xi'-\xi+\mathbf{q}}\varphi^{\xi'}_\mathbf{q-\beta k}\right)
\end{equation}
where $g^{c/v \nu}_\mathbf{q}$ denotes the electron-phonon coupling in conduction and valence band for the phonon mode $\nu$. Furthermore, $\beta=\frac{m_e}{m_e+m_h}$ and $\alpha=\frac{m_h}{m_e+m_h}$ stand for the relative electron and hole masses.
Our ab initio calculations reveal that the electron-phonon matrix elements, similar to the phonon dispersion, only show minor changes in presence of strain allowing us to focus on the unstrained electron-phonon coupling reported in Ref.\cite{Zhenghe2014}. 

%
%
%
\begin{figure}[t!]
\includegraphics[width= \linewidth]{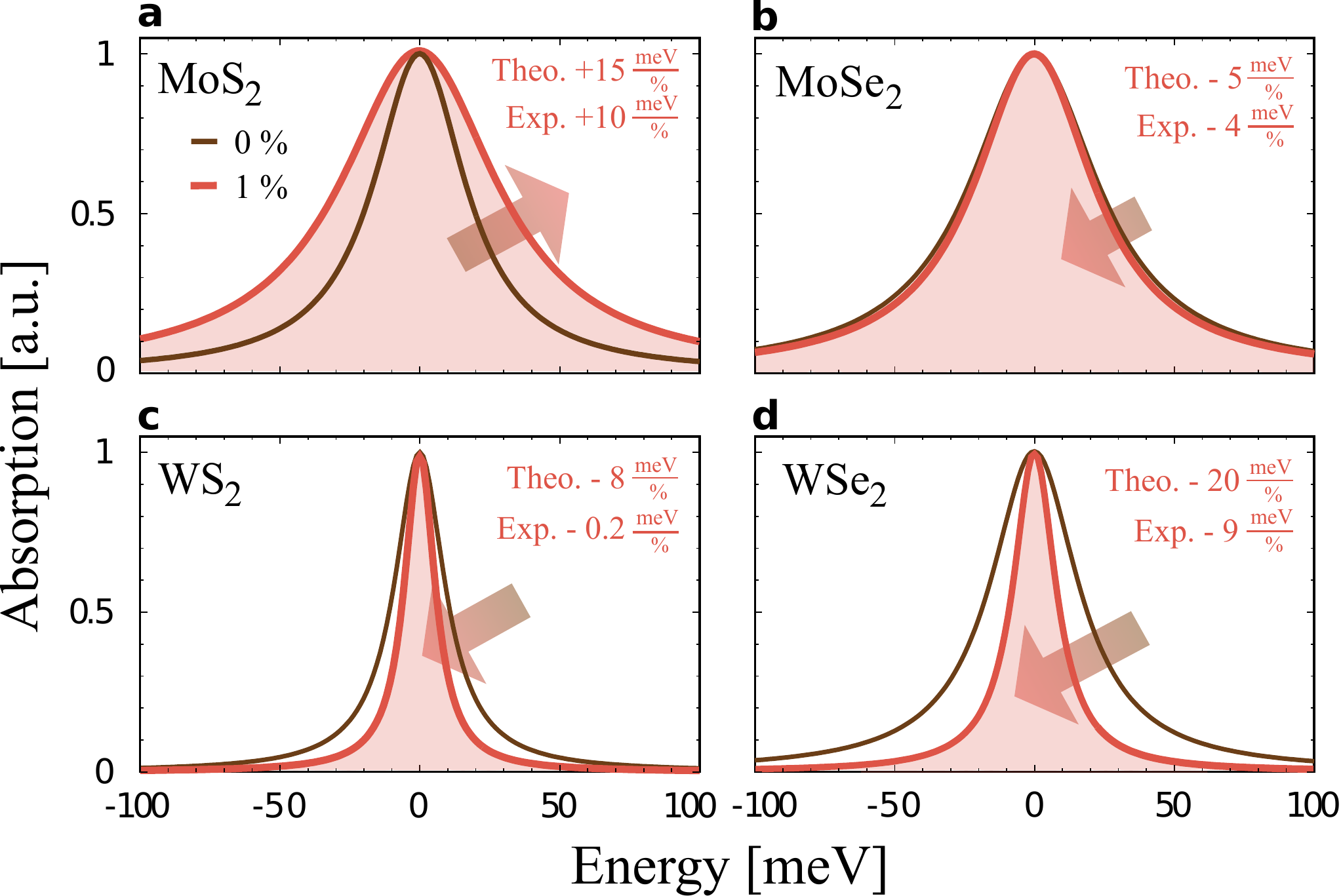}
\caption{Calculated excitonic absorption spectra for four different TMDs with (red) and without (brown) application of $1\%$ tensile strain at room temperature.
The position of the peaks in the absorption spectrum is held fixed at $0~{\rm meV}$ to focus on linewidth changes.
The obtained values are compared with a recent experimental study \cite {Iris2018}. Interestingly, the excitonic linewidth decreases for all TMDs except for ${\rm MoS_2}$, where a clear broadening is theoretically predicted and experimentally observed.
\label{fig3}}
\end{figure}

Finally, the phonon-induced homogeneous linewidth (full width at half maximum) in the absorption spectrum reads \cite{Malte2016}
\begin{equation}
\gamma^{\xi}_\mathbf{Q}=2 \pi \sum_{\mathbf{q},\xi',\nu,\pm} |g^{ \xi\xi'\nu}_\mathbf{q}|^2 \left(\frac{1}{2}\pm \frac{1}{2}+n_\mathbf{q}^{\nu}\right) \delta (E^\xi_\mathbf{Q}-E^{\xi'}_\mathbf{Q+q} \pm \hbar \omega^{\nu}_{\xi'-\xi+\mathbf{q}}).
\label{linewidth}
\end{equation}
Here, the $\delta$ function ensures the energy conservation during the phonon scattering event. 
The $\pm$ sum accounts for the fact that we take phonon emission (+) and absorption (-) processes into account. Furthermore, $\omega^{\nu}_\mathbf{q}$ and $n_\mathbf{q}^{\nu}$ denote the phonon frequency and occupation number of the mode $\nu$ and momentum $\mathbf{q}$, respectively. 
Here, we take explicitly into account acoustic and optical intra- and intervalley phonons, while assuming a phonon bath with constant occupation numbers given by Bose distributions. 
Note that we do not consider the formation of asymmetric linewidths due to phonon-assisted absorption processes \cite{Dominik2017}.

According to Eq. \eqref{linewidth}, it is the excitonic dispersion, the phonon energies and the exciton-phonon coupling matrix elements that determine the excitonic linewidth. 
The rather small changes of the phonon dispersion and the exciton-phonon coupling in presence of strain suggest that the dominant contribution for the experimentally observed strain-induced changes of the linewidth stems from the excitonic dispersion. We can further trace back the strongest impact of strain to the considerable changes in the electronic band structure, while the effect of strain on excitonic properties, such as binding energy and wave functions, is rather small giving rise to changes in the range of 1-5 meV per percent of applied strain \cite{Maja2017}.
\section{Results}
%
%
%
%
%
\begin{figure}[t!]
\includegraphics[width= 0.85\linewidth]{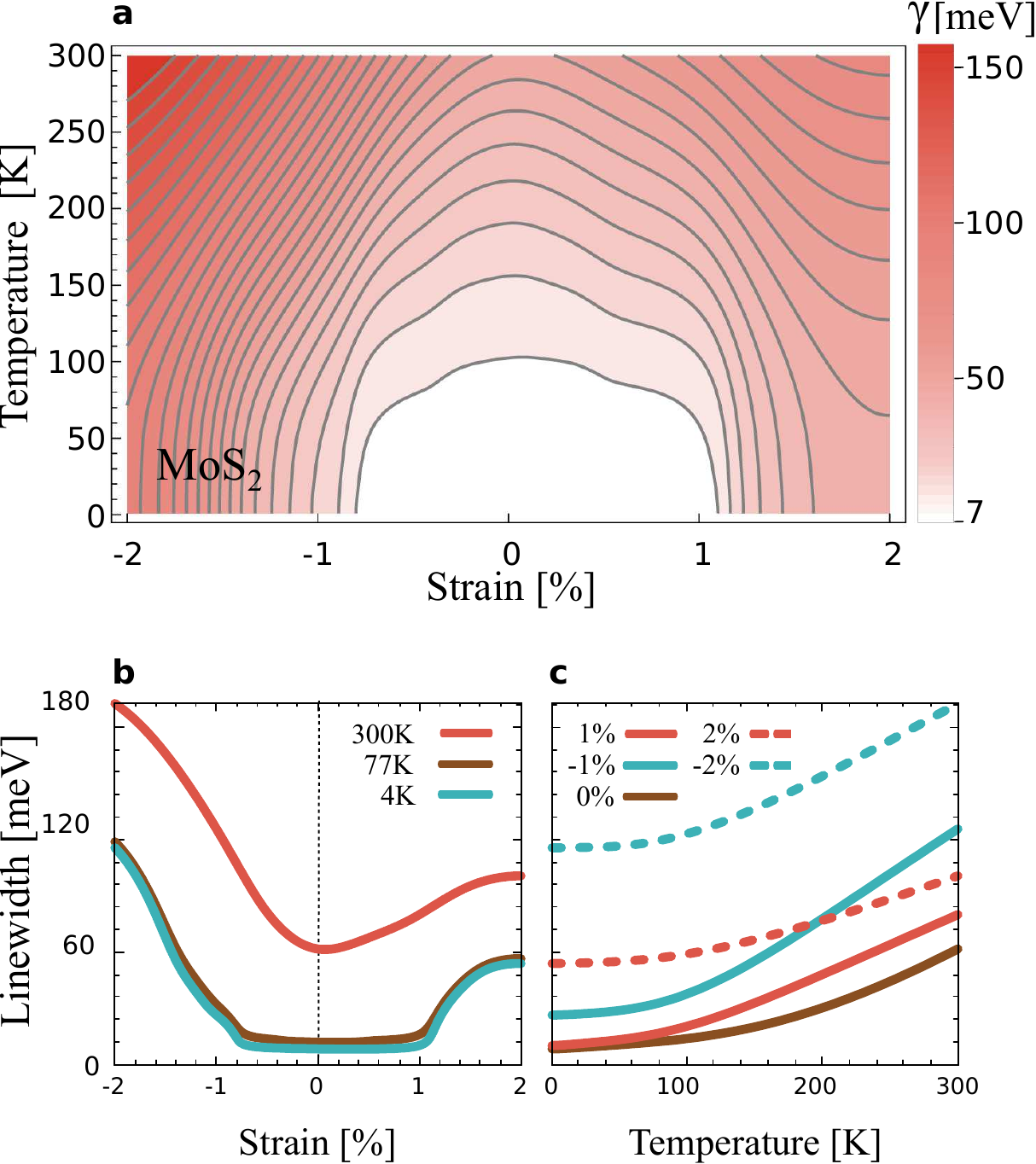}
\caption{(a) Surface plots of the excitonic linewidth as a function of applied tensile and compressive strain and temperature for ${\rm MoS_2}$ monolayer.
Cuts at fixed (b) temperature and (c) strain, respectively.
The smallest linewidth of 7 meV is found for the unstrained case at 0 K, while the largest broadening of 160 meV is found for $-2\%$ compressive strain at room temperature.
\label{fig4}}
\end{figure}

Now, we calculate the excitonic absorption spectrum and provide microscopic insights into the impact of strain.
Figure \ref{fig3} shows the spectra for the four most studied TMD monolayer materials under 1\% tensile strain at room temperature.
Surprisingly, we find that despite the analogous behavior of strain-dependent modification of the electronic dispersion for all TMDs, the behavior of  ${\rm MoS_2}$ is qualitatively different.
While all other TMDs display a decrease in the linewidth in presence of tensile strain,
our calculation reveals a broadening of 15 meV per percent of applied strain for MoS$_2$. This prediction agrees very well with the recent experimentally measured broadening of 10 $\frac{\text{meV}}{\%}$ \cite{Iris2018}.
For MoSe$_2$, WS$_2$, and WSe$_2$, we find a narrowing of the linewidth with the rates of  5, 8, and 20 $\frac{\text{meV}}{\%}$, respectively - again in good agreement with the experimentally observed trends (cf. insets in Fig. \ref{fig3}).

\begin{figure}[t!]
\includegraphics[width= 0.85\linewidth]{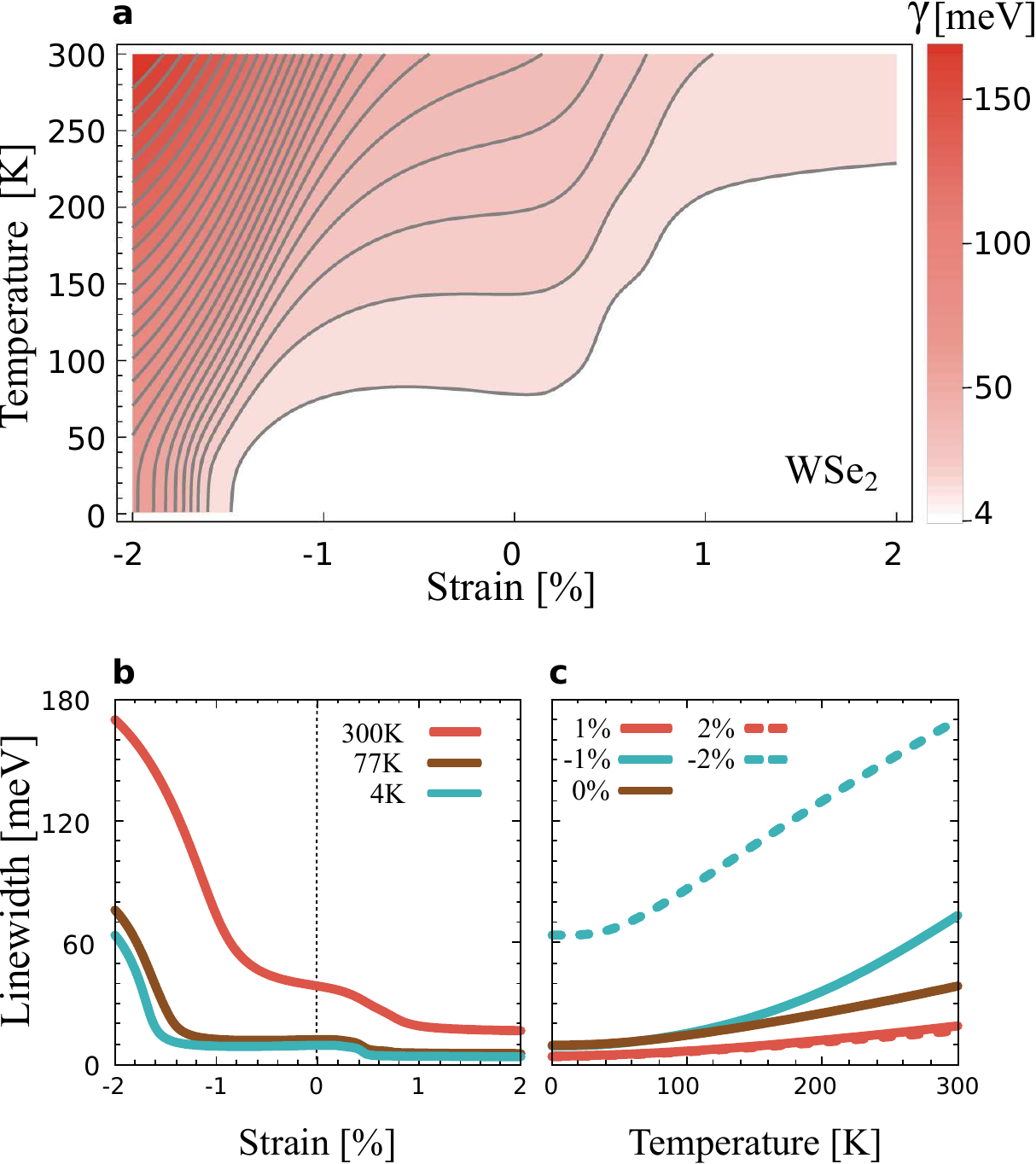}
\caption{Same plots as in figure \ref{fig4} for ${\rm WSe_2}$ monolayer. Now, a qualitatively different behavior is observed: While the excitonic linewidth increases in presence of compressive strain, a clear decrease is found in the case of tensile strain.
\label{fig5}}
\end{figure}

To get a deeper understanding, we show a detailed strain and temperature dependence of the excitonic linewidth for two representative TMD materials (MoS$_2$ and WSe$_2$) in Figs. \ref{fig4} and \ref{fig5}.
Our study includes temperatures from 4 K up to 300 K as well as tensile and compressive strain up to 2$\%$.
We clearly see that the distinct optical response of the TMDs at $1\%$ tensile strain discussed above cannot be generalized to all tensile and compressive strain strengths.
In spite of the linear dependence of the strain-induced changes in the electronic dispersion, the excitonic linewidth is modified non-linearly with respect to the applied strain, which is due to the non-trivial energy and momentum dependence in the scattering integral, cf. Eq. \eqref{linewidth}.
While Figs. \ref{fig4}(a) and \ref{fig5}(a) show surface plots illustrating the dependence of the linewidth on both strain and temperature, the figure parts (b) and (c) represent horizontal and vertical cuts at experimentally most relevant strain and temperature values. 
The most striking difference in the strain dependence of the linewidth between the two materials is that for ${\rm MoS_2}$ the linewidth increases in both strain directions (Fig. \ref{fig4}), whereas the linewidth in ${\rm WSe_2}$ exhibits a monotonous behavior (Fig. \ref{fig5}).
However, for both materials we find a strong increase when applying compressive strain, reaching about four times larger values than in the unstrained case.

%
%

The general increase of the excitonic linewidth with  temperature (Figs. \ref{fig4}(c) and \ref{fig5}(c)) stems from the increased population of phonons $n_\mathbf{q}^{\nu}$ in the system, leading to an enhanced exciton-phonon scattering, cf. Eq. \eqref{linewidth}.
Due to their linear dispersion, scattering with long-range acoustic phonons yields an almost linear temperature dependence, whereas scattering with long-range optical phonons and zone-edge phonons leads to a superlinear increase of the linewidth. Note that possible phonon emission processes ($\propto (1+n_\mathbf{q}^{\nu}$)) give an additional offset of the linewidth since they also occur in the absence of phonons at 0 K.
The temperature dependence of the linewidth is a result of different weights of emission and absorption of acoustic and optical phonons, which change in presence of strain. 
%
%
%
\begin{figure}[!t]
\includegraphics[width= \linewidth]{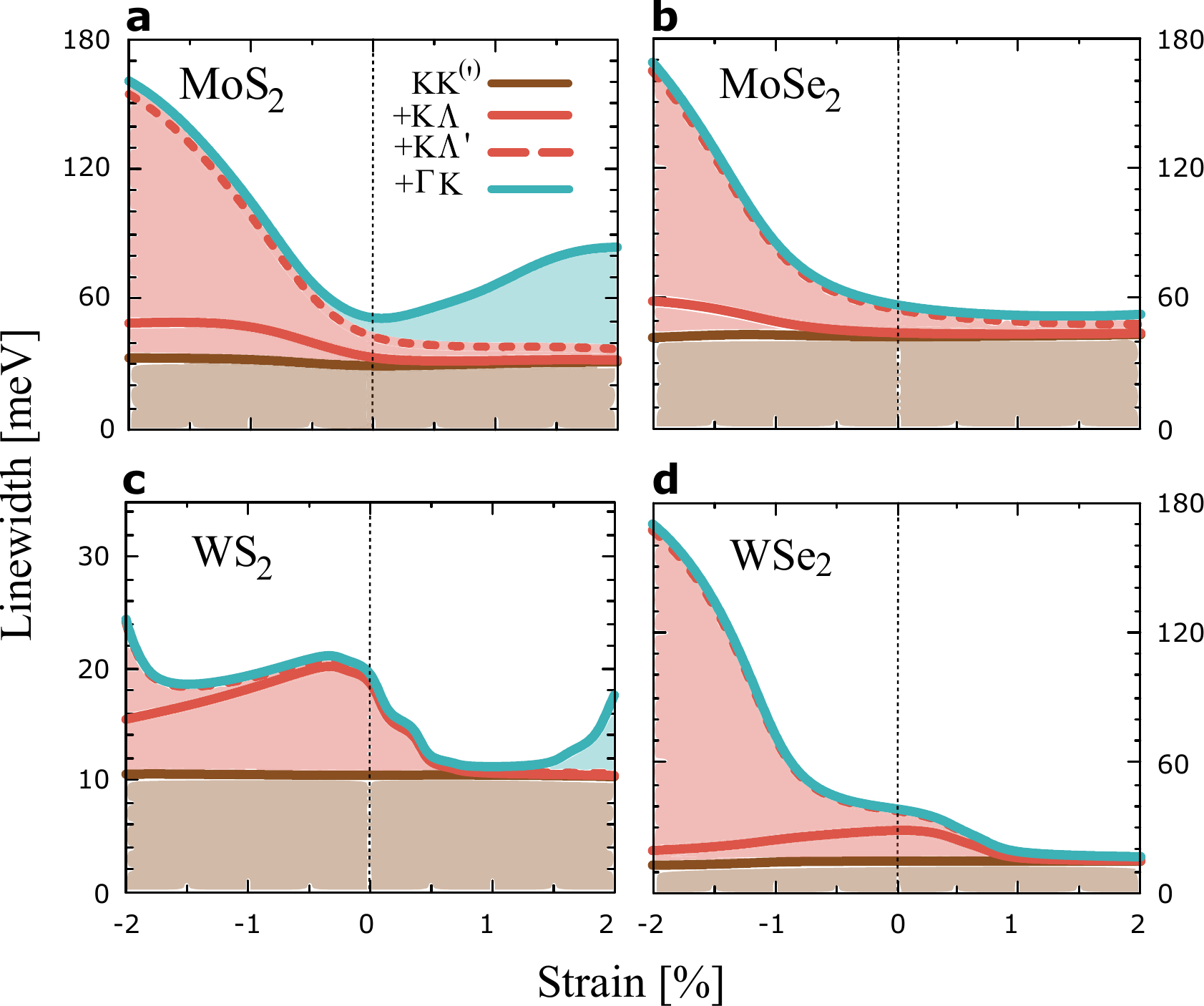}
\caption{Resolution of different exciton scattering channels determining the exciton linewidth for all four investigated TMDs at room temperature.
The shaded brown area denotes the radiative part alongside the KK and the ${\rm KK}^\prime$ scattering. 
The other dominant contribution stems from exciton-phonon scattering into  K$\Lambda$ and K$\Lambda^\prime$ excitons (solid and dashed red) and  $\Gamma$K excitons (blue). Note that the different contributions have been graphically added up successively to pile up to the total linewidth.
All TMDs demonstrate a remarkable broadening of the exciton linewidth in presence of compressive strain as a consequence of the spectral downshift of the K$\Lambda^{(\prime)}$ dark exciton.
\label{fig6}}
\end{figure}

\begin{figure}[!t]
\includegraphics[width= \linewidth]{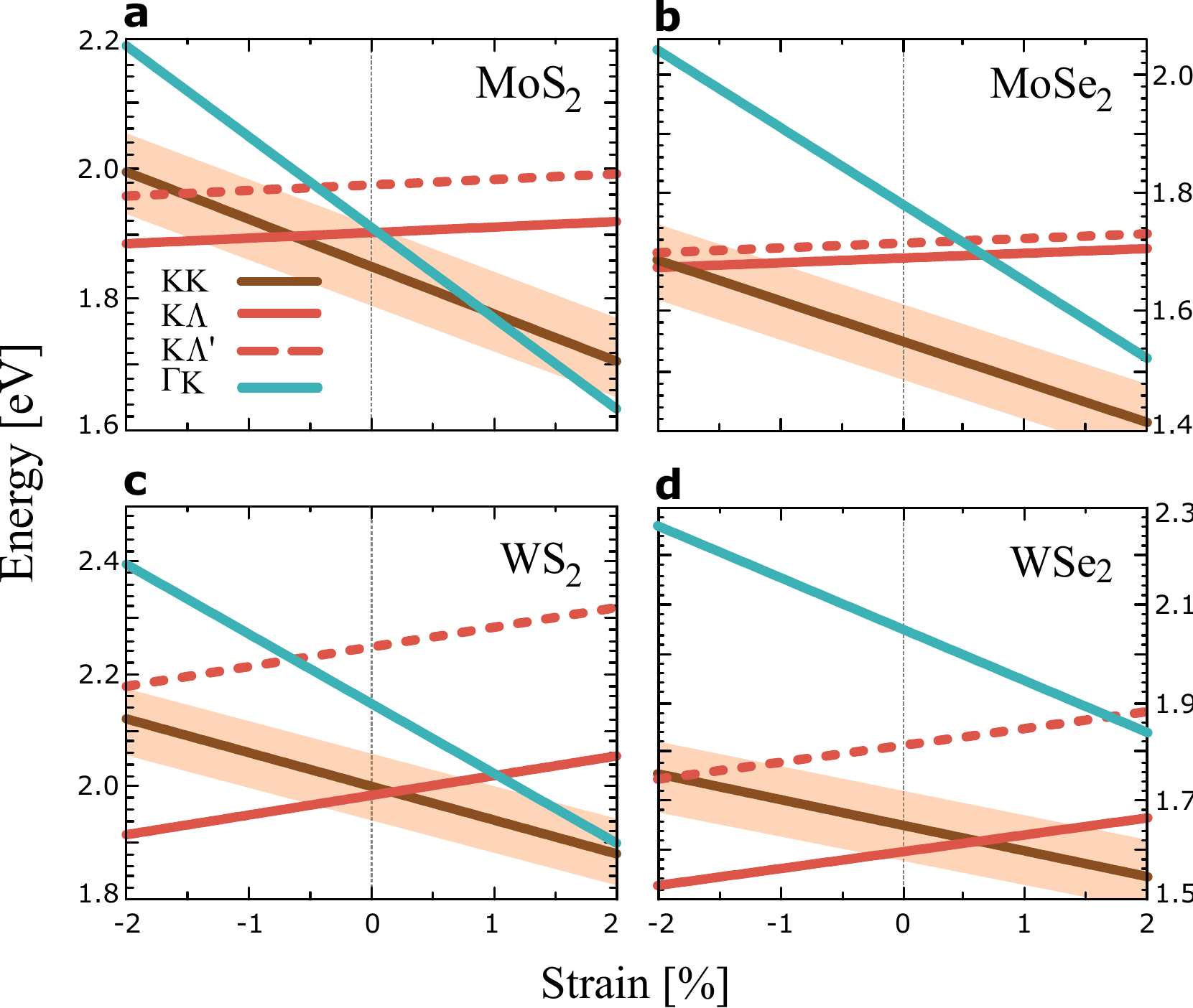}
\caption{Strain-dependent energies of the bright KK and dark K$\Lambda^{(\prime)}$ and $\Gamma$K excitons for all four TMDs.
The shaded orange area describes the energy region around the bright KK exciton that is accessible via phonon absorption/emission contributing to the exciton linewidth. The $\Gamma$K excitons are affected stronger than the K$\Lambda^{(\prime)}$ excitons, since the effective mass at the $\Gamma$ point are modified more efficiently in presence of strain, cf. Table \ref{table1}.
The K$\Lambda$ state is energetically lowest in tungsten-based TMDs in the unstrained case.
In MoS$_2$, the $\Gamma$K state becomes energetically lowest for tensile strain values larger than  $1\%$.
\label{fig7}}
\end{figure}

In order to address this complex interplay of different scattering channels and especially, how it changes with regard to the applied strain, we show a closer look at intra- and intervalley scattering processes behind the excitonic linewidth.
To this end, we divide the sum over all valleys in Eq. (\ref{linewidth}) into single contributions. 
Figure \ref{fig6} illustrates the contribution of the radiative decay and the most dominant phonon-assisted scattering channels including $\rm{KK}$ intravalley processes as well as $\rm{KK}^{\prime}$, K$\Lambda^{(\prime)}$ and $\Gamma$K intervalley scattering channels as a function of strain at room temperature.
In the unstrained case (vertical lines in Fig. \ref{fig6}), the linewidth in Mo-based TMDs is dominated by KK intravalley scattering, whereas for W-based TMDs the intervalley scattering to the K$\Lambda$ dark exciton significantly contributes to the total linewidth. 
For the explanation, it is of crucial importance to have access to the strain-induced modification of the relative position of bright and dark excitonic states. Figure \ref{fig7} shows the energies of KK, K$\Lambda^{(\prime)}$ and $\Gamma$K excitons as a function of strain. The shaded orange area depicts the energy region around the bright KK exciton that is accessible via phonon absorption/emission and thus contributing to the exciton linewidth.
As can be seen from Fig. \ref{fig7}, in unstrained W$-$based TMDs, the K$\Lambda$ dark excitonic state lies below the bright KK state, and thus, is energetically more favorable for scattering with phonons.
Hence, in unstrained W-based TMDs the intervalley K$\Lambda$ dark state contributes dominantly to the total linewidth. 
A detailed discussion on the contribution of diverse dark excitonic states and the coupling strength of relevant phonon modes has been previously studied for the unstrained case \cite{Malte2016}.

%
%
%
%
The intravalley scattering is mostly driven by the absorption of low-energy acoustic phonons and is only sensitive to the modification of the exciton mass determining the required phonon momentum for a scattering process as well as the density of accessible exciton states. However, since the effective masses at the K point only weakly change with strain (less than 2.5\% of their unstrained value per percent strain, cf. Table \ref{table1}), the intravalley scattering also remains almost constant (brown lines in figure \ref{fig6}).
This behavior can be observed well in Fig. \ref{fig6}(b) for ${\rm MoSe_2}$ at room temperature in the strain interval of approximately 1\% to 2\%, where both momentum-forbidden dark K$\Lambda$ and $\Gamma$K states are far away (Fig. \ref{fig7}(b)) and thus only intravalley KK scattering contributes to the linewidth. Therefore, due to the small strain-induced change of the intravalley KK scattering channel, the linewidth stays almost constant.

In contrast, the changes in the intervalley scattering present the dominant mechanism for the observed linewidth variation in presence of strain. 
The smallest modification of the intervalley energy change is $11\%$ per percent strain ($\Gamma$K exciton in ${\rm WSe_2}$, cf. Table.\ref{table1}). Thus, intervalley separations are affected by tensile strain at least $4$ times stronger than excitonic masses.
As an example, we discuss the case of ${\rm WS_2}$ at 300 K, cf. Fig. \ref{fig6}(c). 
At 1\% tensile strain, the relevant scattering partners (K$\Lambda^{(\prime)}$ and $\Gamma$K excitons) are energetically higher than the bright KK state, and hence they are out of the reach for the highest phonon energies, cf. Fig. \ref{fig7}(c). 
At $2\%$ tensile strain, the K$\Lambda$ state shifts even further away, reducing the efficiency of exciton-phonon scattering. However, the $\Gamma$K exciton moves closer to the KK state and thus leads to an enhanced scattering via absorption of phonons.
Overall, the excitonic linewidth becomes almost twice broader than its value at 1\% tensile strain, cf. Fig. \ref{fig6}(c).

As already seen in Fig. \ref{fig5}, ${\rm WSe_2}$ shows a qualitatively different behavior in presence of tensile strain. The excitonic linewidth decreases monotonously, when tensile strain is applied (Fig. \ref{fig6}(d)), which can be traced back to a strain-induced quenching of exciton-phonon scattering channels.
The K$\Lambda$ exciton, which is the energetically lowest state in the unstrained ${\rm WSe_2}$ monolayer, shifts up with the rate of 87 $\frac{{\rm meV}}{\%}$ (Fig. \ref{fig7}(d)). This drastically reduces the exciton-phonon scattering into this state. Despite the fact that the $\Gamma$K state moves closer to the KK exciton, it still stays beyond the energetically accessible region by phonons, at least in the investigated range of strain values, cf. the orange shaded region in Fig. \ref{fig7}(d). 
Consequently, the exciton linewidth decreases with the tensile strain in ${\rm WSe_2}$.
In contrast, MoS$_2$ resembles more the case of WS$_2$ when being stretched, as here the $\Gamma$K exciton moves closer to the KK state (Fig. \ref{fig7}(c)) and thus, the linewidth broadens. This occurs first via absorption of phonons and at strain values larger than $1.5\%$ even emission becomes possible, since the K$\Gamma$ exciton shifts below the bright state.

To guide future experimental studies, we now, discuss the case of compressive strain and low temperature.
The behavior of the excitonic linewidth for all investigated TMDs in presence of compressive strain is determined by the scattering to the K$\Lambda^{(\prime)}$ state. They shift to lower energies, while the $\Gamma$K state moves energetically far away, cf. Fig. \ref{fig7}.
As a result, the linewidth increases as a consequence of the scattering into the K$\Lambda^{(\prime)}$ states, cf. Fig. \ref{fig6}.
Furthermore, for all TMDs except for WS$_2$, K$\Lambda^{\prime}$ excitons contribute more efficiently to the excitonic linewidth compared to the K$\Lambda$ states in presence of compressive strain cf. Fig. \ref{fig6}.
This is due to the fact that K$\Lambda^{\prime}$ excitons have 5 times stronger electron-phonon coupling than that of K$\Lambda$ states.
For WS$_2$, the K$\Lambda^{\prime}$ exciton is so distant that it contributes to linewidth broadening only at 2\% compressive strain, where it falls in the accessible phonon region (Fig. \ref{fig7}).
In general, the influence of strain is much more efficient for compressive strain due to the enabled emission processes to the energetically lower lying dark states and the much stronger carrier-phonon interaction for M phonons enabling the scattering into the K$\Lambda^{\prime}$ state \cite{Zhenghe2014}.

For low temperatures, we study the case of ${\rm MoS_2}$ and ${\rm WSe_2}$, cf. Fig. \ref{fig4}(b) and Fig. \ref{fig5}(b). For ${\rm MoS_2}$ in the strain interval of approximately -0.65\% to 0.9\% both momentum-forbidden dark K$\Lambda$ and $\Gamma$K states are higher than the KK state (Fig. \ref{fig7}(a)). Here, the broadening due to intravalley absorption of acoustic phonons is small, so that the changes with strain purely result from the modification of intervalley scattering channels. Therefore, since the majority of the phonon population is accessible only through emission at low temperature, the linewidth is only determined by the radiative decay, cf. Fig. \ref{fig4}(b). 
Above 0.65\% compressive (0.9\% tensile) strain, the K$\Lambda$ ($\Gamma$K) state becomes energetically lower than the KK exciton and leads to broader linewidth. As discussed before, the broadening due to compressive strain is more pronounced compared to tensile strain.
For ${\rm WSe_2}$, the only intervalley scattering channel is the K$\Lambda^{(\prime)}$ exciton that is located energetically lower than the KK state for strain values smaller than 0.3\% and all compressive strain values. Thus, we observe linewidth broadening, when compressing ${\rm WSe_2}$ monolayer starting at 0.3\% uniaxially strained ${\rm WSe_2}$. Moreover, due to lack of phonon absorption at low temperatures,  the linewidth is only determined by radiative decay for larger tensile strain values (Fig. \ref{fig6}(d)).

\section{Conclusion}
We have performed a microscopic study on the impacts of strain on TMDs combining the semiconductor Bloch equations and first-principle calculations. We demonstrate that the unexpectedly different strain-dependent changes in the excitonic linewidth of diverse TMD monolayers can be traced back to a large extent only to the strain-induced changes in the relative spectral position of bright and dark excitonic states. The latter determines the efficiency of exciton-phonon scattering, which is the dominant mechanism behind the linewidth broadening in the considered low-excitation limit. Our results are in very good agreement with recent experimental studies on the strain-controlled optical response of various TMD materials. The gained insights might guide future experimental studies, in particular in the field of low-temperature and compressive strain measurements. 

\section{Acknowledgment}
This project has received funding from the European Union’s Horizon 2020 research and innovation program under grant agreement No 696656. Furthermore, we acknowledge financial support from the Swedish Research Council (VR) and the Deutsche Forschungsgemeinschaft (DFG) through SFB 951 and the School of Nanophotonics (SFB 787).

\section{References}
\bibliography{ref.bib}

\end{document}